# Efficient Multiple Exciton Generation in Monolayer MoS$_2$

*Ashish Soni [1,2], Dushyant Kushavah [1,2], Li-Syuan Lu [3], Wen-Hao Chang[3,4] and Suman Kalyan Pal [1,2]\**

[1] School of Physical Sciences, Indian Institute of Technology Mandi, Kamand, Mandi 175005 Himachal Pradesh, India

[2] Advanced Materials Research Centre, Indian Institute of Technology Mandi, Kamand, Mandi 175005 Himachal Pradesh, India

[3] Department of Electrophysics, National Yang Ming Chiao Tung University, Hsinchu 30010, Taiwan

[4] Research Center for Applied Sciences, Academia Sinica, Nankang, Taipei 11529, Taiwan

[*] Email: suman@iitmandi.ac.in

## Abstract

Utilizing the excess energy of photoexcitation that is otherwise lost as thermal effects can improve the efficiency of next-generation light-harvesting devices. Multiple exciton generation (MEG) in semiconducting materials yields two or more excitons by absorbing a single high-energy photon, which can break the Shockley-Queisser limit for the conversion efficiency of photovoltaic devices. Recently, monolayer transition metal dichalcogenides (TMDs) have emerged as promising light-harvesting materials because of their high absorption coefficient. Here, we report efficient MEG with low threshold energy and high (86%) efficiency in a van der Waals (vdW) layered material, MoS$_2$. Through different experimental approaches, we demonstrate the signature of exciton multiplication and discuss the possible origin of decisive MEG in monolayer MoS$_2$. Our results reveal that vdW-layered materials could be a potential candidate for developing mechanically flexible and highly efficient next generation solar cells and photodetectors.

**Introduction**

Transition metal dichalcogenides (TMDs) possess a layered structure with weak interlayer interactions and hence can be thinned downed to one atomic layer. The atomically thin TMDs exhibit many fascinating properties including a wide range of optical bandgaps, mechanical flexibility, high absorption coefficient (one order of magnitude higher than conventional semiconductors), near-ideal optical bandgap for solar energy harvesting, and self-passivated surface.[1-5] These intriguing optoelectronic properties of two-dimensional (2D) TMDs make them promising candidates for flexible photodetectors and photovoltaics[6] where light converts into electron-hole pairs. A detailed balance calculation predicted that the power conversion efficiency (PCE) of single-junction solar cells of 2D TMD can reach up to 27%.[7] But, the PCE achieved till date in thin-film single-junction TMD solar cells is relatively low (~5-6%).[8, 9]

MEG is a process where additional electron-hole pairs (excitons) are generated from a high-energy absorbed photon. A photon of energy ($hv$) higher than the optical bandgap ($E_g$) can excite the ground state population to a higher energy band and create hot electron-hole pairs in a semiconducting material. Usually, these high-energy electron-hole pairs cool down to the band-edge by losing their excess energy ($hv$-$E_g$) in the form of heat. However, in the process of MEG, a high energy charge carrier (electron in the conduction band or hole in the valance band) relaxes to the band-edge by lifting an electron from the valance band to the conduction band and thus producing a new exciton. Therefore, the efficiency of conversion of light into charge pairs can be greatly enhanced when multiple excitons are produced by hot carriers without losing their excess energy as heat. The MEG process can be characterized by the quantum yield (QY) (number of excitons generated per absorbed photon) and the threshold photon energy, i.e., the energy at which MEG starts. If there are 2(n) excitons generated from a single absorbed photon, then QY is said to be 2(n). In the ideal case, QY reaches the value 2(n) at a photon energy twice (n times) the optical bandgap energy. In bulk semiconductors, carrier multiplication is known as Impact Ionization (II), whereas it is called MEG in quantum-confined structures. The efficiency of MEG is very low in the bulk semiconductors because of the strong electron-phonon scattering and high screening of coulombic interactions.[10, 11] The MEG threshold in bulk materials is almost four times the optical bandgap energy.[12] On the contrary, MEG has been seen to be improved in the low dimensional material as the carrier relaxation rate through electron-phonon interactions can be reduced because of the discrete energy of the charge carriers due to strong confinement in the reduced dimension.[13, 14]

In 2002, Nozik proposed that MEG is more efficient in nanomaterials than their bulk counterpart.[15] Later, MEG has been demonstrated in many colloidal quantum dots, nanorods, carbon nanotubes, and graphene.[16, 17] Because of the strong quantum confinement arising from their reduced dimensionality, 2D materials could be an interesting platform to study MEG. Indeed, Zhou et al.[18] investigated MEG in few-layer black phosphorus (BP). Though, MEG parameters are very promising, BP has high chemical reactivity, which leads to quick degradation of BP in ambient conditions.[19, 20] Kim and colleagues recently observed efficient MEG in TMD ($MoTe_2$ and $WSe_2$) films.[21] Among the family of layered TMDs, $MoS_2$ is one of the most investigated materials. It has been proven to be a promising material for low-cost, scalable, and novel optoelectronic devices.[22-24] Moreover, monolayer $MoS_2$ and BP can form a p-n heterojunction of type II, which could generate free carriers via exciton dissociation and hence could be useful for efficient extraction of multiple carriers either from multiple excitons formed in BP or $MoS_2$. However, possibility of photoinduced carrier multiplication in mono- or multilayer $MoS_2$ remains unexplored. The monolayer TMDs could be beneficial for realizing more compact and flexible optoelectronic devices for the future. Although generation of multiple excitons was proposed in multilayer TMDs, the mechanism of carrier multiplication and the direct spectroscopic evidence of this higher excitonic process are still elusive.

Herein, we demonstrate MEG in atomically thin $MoS_2$ using ultrafast transient absorption spectroscopy. The monolayer sample was excited with photons of energies higher than the optical bandgap energy (i.e., $E_{ex} > E_g$). The photoinduced bleach (PIB) signal measured at an excitation energy greater than $2E_g$ was found to be much stronger than that of the signal for band edge excitation suggesting MEG in $MoS_2$ sample. The delayed rise of PIB signal for excitation energy greater than $2E_g$ further supports carrier multiplication which occurs via inverse Auger process within 160 fs. We report enhanced MEG with threshold down to $2.1E_g$ and slope efficiency up to 86% in monolayer $MoS_2$.

**Results and discussion**

**Monolayer $MoS_2$.** Atomically thin $MoS_2$ flakes were prepared on a double-polished optically transparent sapphire substrate via chemical vapor deposition (CVD). Optical microscopy image in figure 1(a) confirms formation of triangular flakes of $MoS_2$ which spread over the whole substrate. The linear absorption spectrum of the sample measured at room temperature is presented in figure 1(b), which depicts three distinct peaks centered at 1.88 eV (660 nm), 2.03 eV (610 nm), and 2.88 eV (430 nm).[25-27] The first two peaks labeled by A and B correspond to excitonic transition at K/K′ point of the Brillouin zone. These two peaks arise as a result of the transitions between splitted valance and conduction bands due to spin-orbit coupling in $MoS_2$.[28-30] The high energy peak, i.e., C with strong absorption, is associated with the transitions involving the bands between the Γ-K point of the Brillouin zone[31, 32] and has been ascribed to the van Hove singularities in density of state (DOS).[33] It is worth noting that the absorption spectrum of our prepared sample resembles the spectrum of monolayer $MoS_2$.[30] To confirm the sample quality and estimate the layer number present in the sample, Raman spectroscopy was employed. Similar to the previously reported Raman spectrum,[34] our sample exhibits two characteristic peaks corresponding to $E^1_{2g}$ (in-plane vibration) and $A_{1g}$ (out-of-plane vibration) at 385 cm⁻¹ and 404 cm⁻¹, respectively. The peak shift difference between two Raman peaks is 19 cm⁻¹ and $E^1_{2g}/A_{1g}$=1.04≈1, which are suggesting the monolayer nature of our $MoS_2$ flakes.[34, 35] It should be noted that the third peak (~419 cm⁻¹) in the Raman spectrum is due to the sapphire substrate.[36, 37] For completeness, photoluminescence (PL) spectrum of our sample was measured following excitation by a 2.33 eV laser light. The PL spectrum has a peak at 1.86 eV and it is matching well with the previously reported PL spectrum of monolayer $MoS_2$.[38-40]

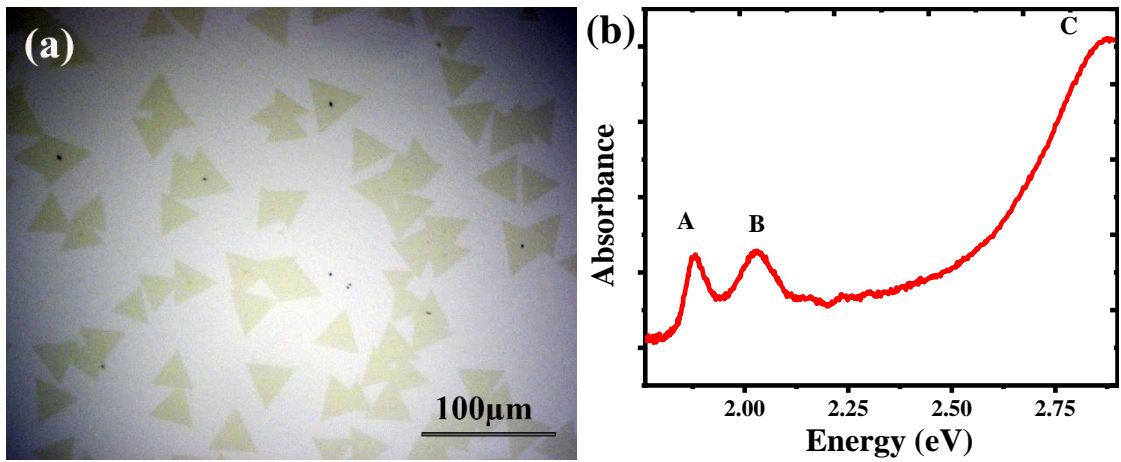

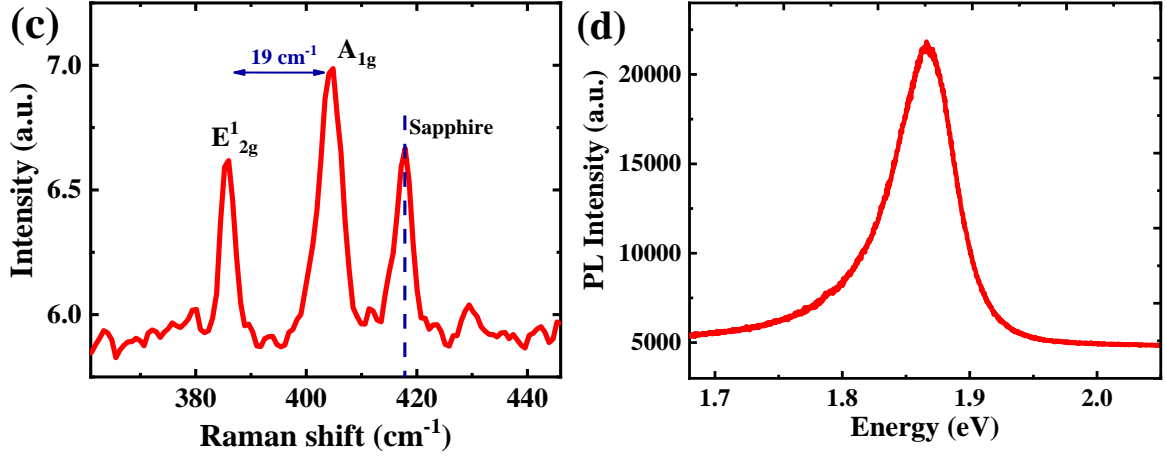

**Figure 1** (a) Optical image of CVD grown monolayer MoS$_2$ on the sapphire substrate. (b) Absorption spectrum of CVD grown MoS$_2$ sample. (c) Raman spectrum of monolayer MoS$_2$ obtained with 532 nm laser light. (d) Photoluminescence spectrum of MoS$_2$ monolayer (excitation photon energy 2.33 eV).

**Photoinduced transient absorption study.** Femtosecond pump-probe technique was employed to monitor the ultrafast dynamics of the optically generated charge carrier in monolayer MoS$_2$. The dependence of transient absorption (TA) of monolayer MoS$_2$ on probe photon energy at different time delays (between pump and probe beams) is presented in figure 2. This spectral measurement was carried out at an excitation energy of 2.33 eV (532 nm) and fluence $3.7 \times 10^{13} cm^{-2}$ at room temperature. It can be seen that the TA spectra are featured with two negative (at 1.82 and 2.0 eV) and two positive bands (~ 1.76 and 1.91 eV). The reason for positive band in TA spectra is increase of photoinduced absorption, whereas negative band arises due to decrease of photoinduced absorption of the probe.[41, 42] After excitation, the photogenerated charge carriers undergo thermalization followed by phase space filling resulting decrease of absorption of time-delayed probe as

$$\Delta A = A_{pump\ on} - A_{pump\ off} \qquad 1$$

where A is absorbance. The absorption peaks at 1.88 and 2.03 eV in the steady state absorption spectrum of monolayer $MoS_2$ (figure 1b) corresponds to A and B excitons, respectively. Therefore, the negative features at 1.82 and 2.0 eV could be ascribed to photoinduced bleach (PIB) due to A and B excitonic transitions, respectively.[43] The positive peaks around 1.76 and 1.91 eV could arise due to a number of processes, viz. the broadening of the main bleach feature, bandgap renormalization or photoinduced absorption (PIA) of excited carriers. However, these features could also be due to biexciton formation as reported by previous literature.[43-46]

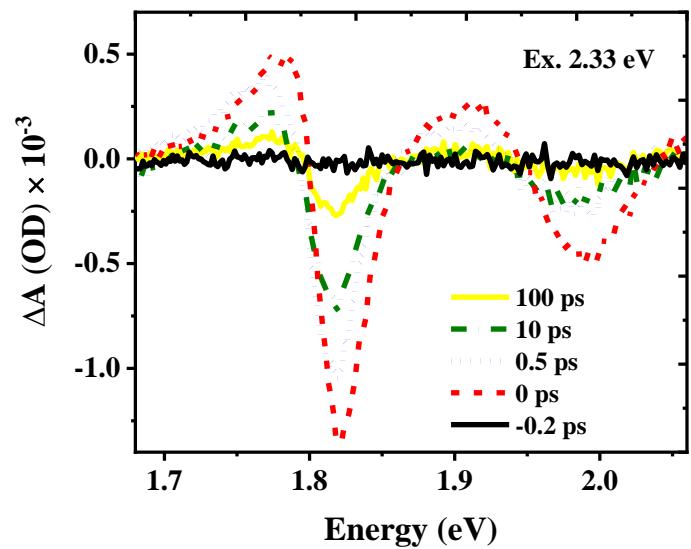

**Figure 2** Transient absorption spectra for monolayer $MoS_2$ measured at a pump photon energy 2.33 eV and pump fluence $3.7 \times 10^{13} cm^{-2}$.

**Decay kinetics of photoexcited carriers.** Transient absorption (TA) spectroscopy has been proven to be a very successful technique to study the MEG in quantum-confined semiconductor nanostructures,[47-49] Previous studies on MEG in quantum dots reported abrupt change in the decay kinetics of the PIB signal in changing the excitation energy from below to above the threshold energy of MEG.[12, 13, 50, 51] The change in PIB kinetics was attributed to the Auger recombination of the generated multiple excitons within the same quantum dot.

Therefore, it is crucial to investigate the carrier dynamics to check the possibility of MEG in a material. First, we recorded PIB signal at 1.86 eV for different pump fluence in monolayer MoS$_2$ at an excitation photon energy of 2.33 eV (1.20 E$_g$), where MEG is not possible (figure 3(a)). The variation of maximum absorption of PIB signal ($\Delta A_{max}$) with absorbed photon fluence (see supporting information for determination of absorbed photon fluence from incident fluence) was plotted (inset of figure 3(a)) and obtained a straight line up to an absorbed photon fluence of $1.0 \times 10^{14}$ cm$^{-2}$.

Furthermore, the decay kinetics of the PIB signal can be fitted globally with a multi-exponential model as given bellow

$$\Delta A = A_1 \, exp(-t/T_1) + A_2 \, exp(-t/T_2) + \cdots \qquad 2$$

where $T_i$ ($i$=1, 2, 3…) is the decay time, and $A_i$ ($i$=1, 2, 3…) represents the amplitude of the i$^{th}$ decay component. We fitted PIB kinetics using equation 2 and fitting results are presented in the supplementary Table S1. The decay kinetics at lower fluences can be fitted with a bi-exponential function (a fast sub-picosecond and a slow decay component with time constants ~0.2 ps and ~150 ps, respectively), but one extra time component (~13 ps) is required to properly fit the data at high pump fluences. The rapid rise of the bleach signal is due to the phase space filling of charge carriers near the conduction band. Considering a report on large-area monolayer MoS$_2$ by Cunningham,[52] we attribute the fast component of the decay to carrier capture by either trap or mid-gap defect states. The slowest decay of PIB signal could be ascribed to the radiative recombination of excitons.[52, 53] The intermediate time component observed at pump fluences above $7.0 \times 10^{13}$ cm$^{-2}$ possibly arises due to the Auger recombination of excitons as it matches well with the reported time constant for the Auger recombination in MoS$_2$.[54]

Next, we measured the fluence-dependent PIB kinetics for 3.87 eV (2.1 $E_g$) excitation photon energies which can be seen in figure 3(b). The maximum PIB signal varies linearly with absorbed photon fluence till an absorbed photon fluence of $0.4 \times 10^{14}$ cm$^{-2}$ (inset of figure 3(b)). Interestingly, the PIB kinetics at this pump energy (>2$E_g$) fitted well globally with three time-components: ~0.23 ps, 14 ps, and ~150 ps (fitting results are presented in the supplementary Table S2). The origin of the decay components could be the same as that mentioned before for pump photon energy 2.33 eV.

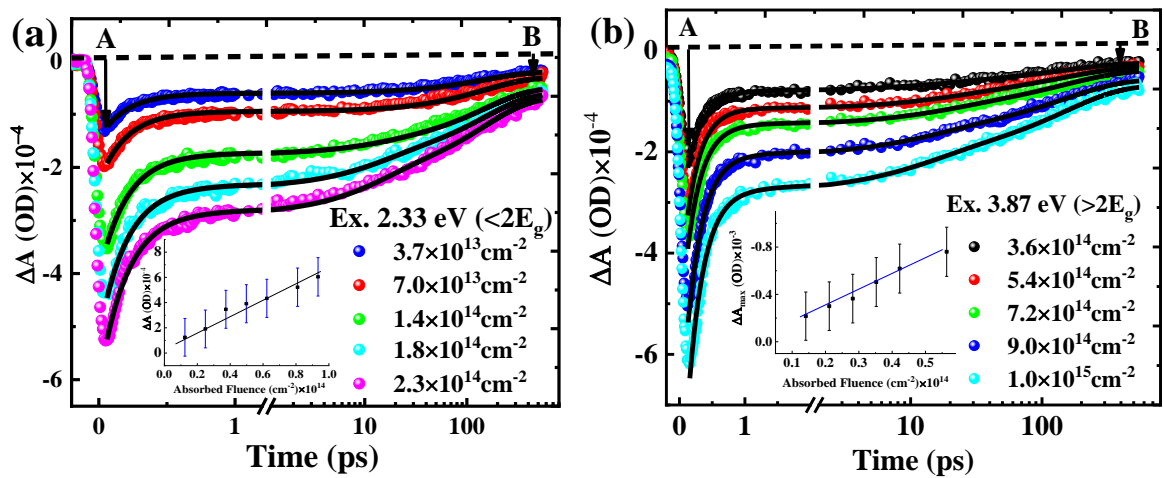

**Figure 3.** PIB kinetics probed at direct A-exciton (1.86 eV) transition using pump photon energy (a) 2.33 eV and (b) 3.87 eV (at different pump fluences). Solid lines are results of global fitting. Inset: variation of maximum PIB signal with absorbed photon fluence.

**Observation of MEG.** We measured PIB kinetics for excitation photon energies 1.94, 2.6, 2.88, and 4.4 eV in addition to 2.33 and 3.87 eV. To determine the linear regime of pump fluence, we plotted $\Delta A_{max}$ with pump photon fluence (supplementary figure S2).[13, 21, 55] The decay kinetics measured at excitation energy 2.33 eV and at different pump fluences are shown in figure 4(a). All the PIB kinetics measured at an excitation energy of 2.33 eV almost collapse on the same curve after normalization (inset of figure 4a). Similar trend has been observed for other excitation photon energies (see supplementary figure S3). These observations infer very similar nature of decay dynamics at all measured fluences and the maximum PIB signal is proportional to the number of generated excitons. Hence, we have compared the density of generated excitons by comparing TA kinetics at different photon energies after normalizing at 200 ps (where signal is due to long-lived single exciton) to detect the MEG,[13, 56, 57] and shown in figure 4(b). It is clear from Figure 4(b) that the maximum PIB signal becomes double as the excitation energy ($E_{ex}$) is increased to more than $2E_g$, i.e., $2E_g < E_{ex} < 3E_g$. However, the inset of figure 4(b) (normalized TA kinetics) indicates that decays follow precisely the same dynamics for all the excitation energies less than $2E_g$. These observations suggest that the number of excitons generated per absorbed photon is more for excitation photon energies >$2E_g$ than that for excitation energies less than $2E_g$.

Next, we have compared the build-up dynamics of the PIB signals probed at A-excitonic transition for 2.33 eV (1.2 $E_g$) and 3.87 eV (2.1 $E_g$) pump energies (figure 4(c)). A quick rise for pump energy <$2E_g$ has been found, which can be attributed to the hot carrier relaxation process via phonon emission (I), whereas a relatively slow rise is observed in the case of pump energy >$2E_g$. The slower rise time of PIB signal for high energy excitation had previously been reported as a possible fingerprint of MEG by Kim et al.[21] Therefore, the slower rise time of PIB signal for excitation photon energy above $2E_g$ provides further evidence of MEG (II) in monolayer $MoS_2$. The processes responsible for the relaxation of hot carriers for excitation photon energy <$2E_g$ and >$2E_g$ are schematically presented in figure 4(d).

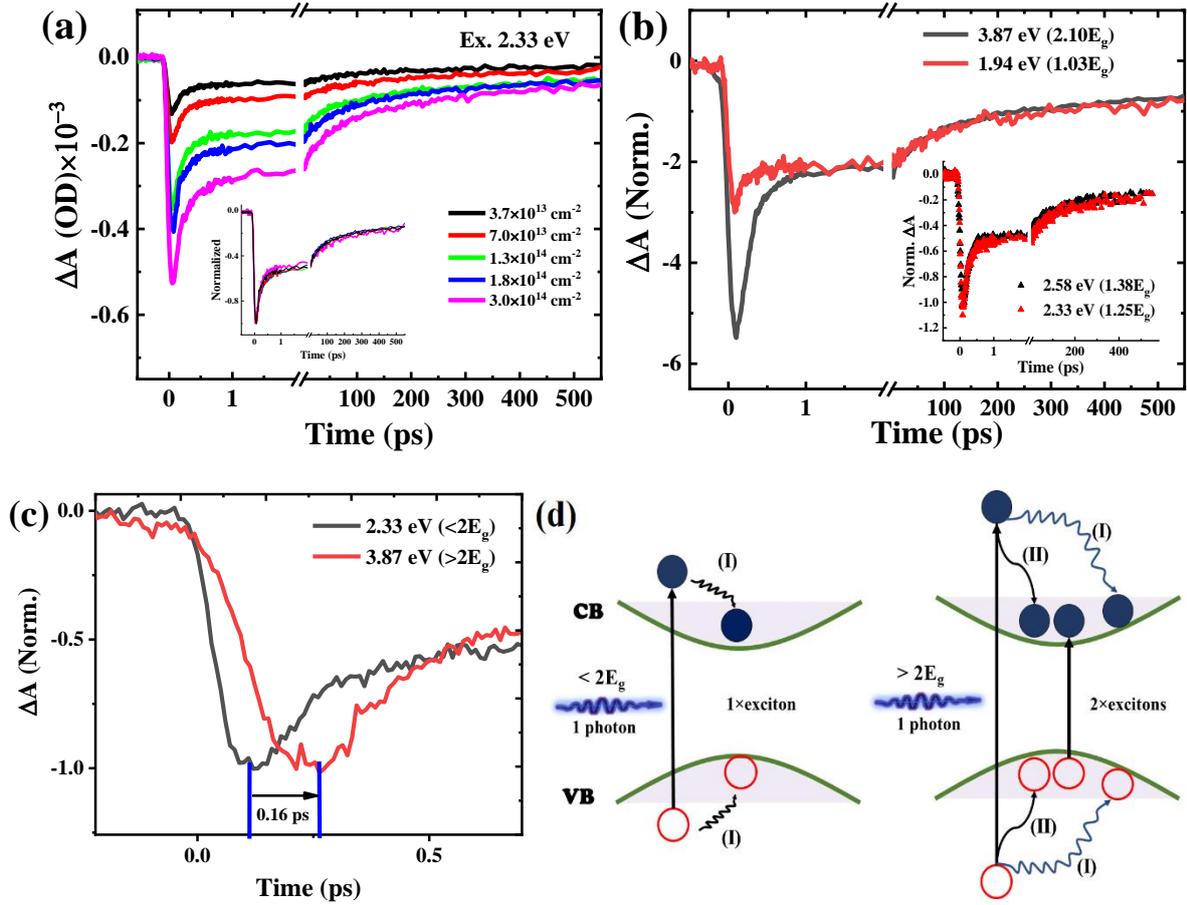

**Figure 4** (a) Kinetics of PIB at different pump fluences measured with photon energy 2.33 eV (Inset: PIB kinetics after normalization). (b) Bleach kinetics at different pump energies (1.94 and 3.87 eV) normalized at a long delay time (Inset: PIB kinetics at two energies (<2$E_g$) after normalization). (c) Build-up dynamics of PIB signal for below and above threshold excitations. (d) Schematic representation of possible relaxation pathways in case of below and above threshold excitations.

Furthermore, we globally fitted bleach signals measured at pump energies 2.33 eV (1.2$E_g$), 3.87 eV (2.1$E_g$) and 4.4 eV (2.4$E_g$) with absorbed photon fluence $2.8 \times 10^{13}$ cm$^{-2}$ using equation 2 (supplementary figure S4). It is evident from the fitting results (supplementary Table S3) that an additional decay component (time constant ~ 17 ps) is present for pump energies >2$E_g$, which corresponds to the Auger recombination of excitons (as mentioned before). The presence of the additional decay channel provides further evidence for the generation of multiple excitons which decay via Auger processes with faster rate as compared to single exciton decay.

The maximum intensity ($\Delta A_{max}$) of PIB signal can be considered as a measure of exciton density if it increases linearly with the absorbed photon fluence. Therefore, to compare the number of exctons generated at different excitation energies, we plotted $\Delta A_{max}$ of the PIB signal at different excitation energies as a function of the absorbed photon fluence (figure 5(a)). It is apparent from figure 5(a) that the maximum PIB intensity increases linearly with the absorbed pump fluence for all pump energies. The maximum PIB signal ($\Delta A_{max}$) depends on the quantum yield ($\emptyset$) of exciton generation and the absorbed pump fluence ($F_{abs}$) as

$$\Delta A_{max} = \emptyset \sigma F_{abs} \qquad\qquad 3$$

where $F_{abs} = fraction\ of\ absobed\ photon \times pump\ fluence$ and $\sigma$ is the transient bleach cross-section for the probe.[13] It follows from equation 3 that the slope of a fitted line in figure 5(a) represents the exciton generation yield. Without multiple carrier generation, the slope should remain constant for all the excitation energies less than $2E_g$ (i.e., 2.33, 2.58, and 2.88 eV), which is indeed the case, i.e., the slope of $\Delta A_{max}$ vs. absorbed fluence is almost same (figure 5(a)). For excitation energies greater than $2E_g$, the slope of the straight-lines are similar for all excitation energies ($E_{ex}$). But, the slope of $\Delta A_{max}$ $vs$ $F_{abs}$ lines for $E_{ex} > 2E_g$ is about twice than that of $E_{ex} < 2E_g$ lines (figure 5(a)). Such abrupt and sudden increase in slope of $\Delta A_{max}$ vs $F_{abs}$ lines at excitation energies above $2E_g$ could be associated to multiple exciton generation in monolayer $MoS_2$.

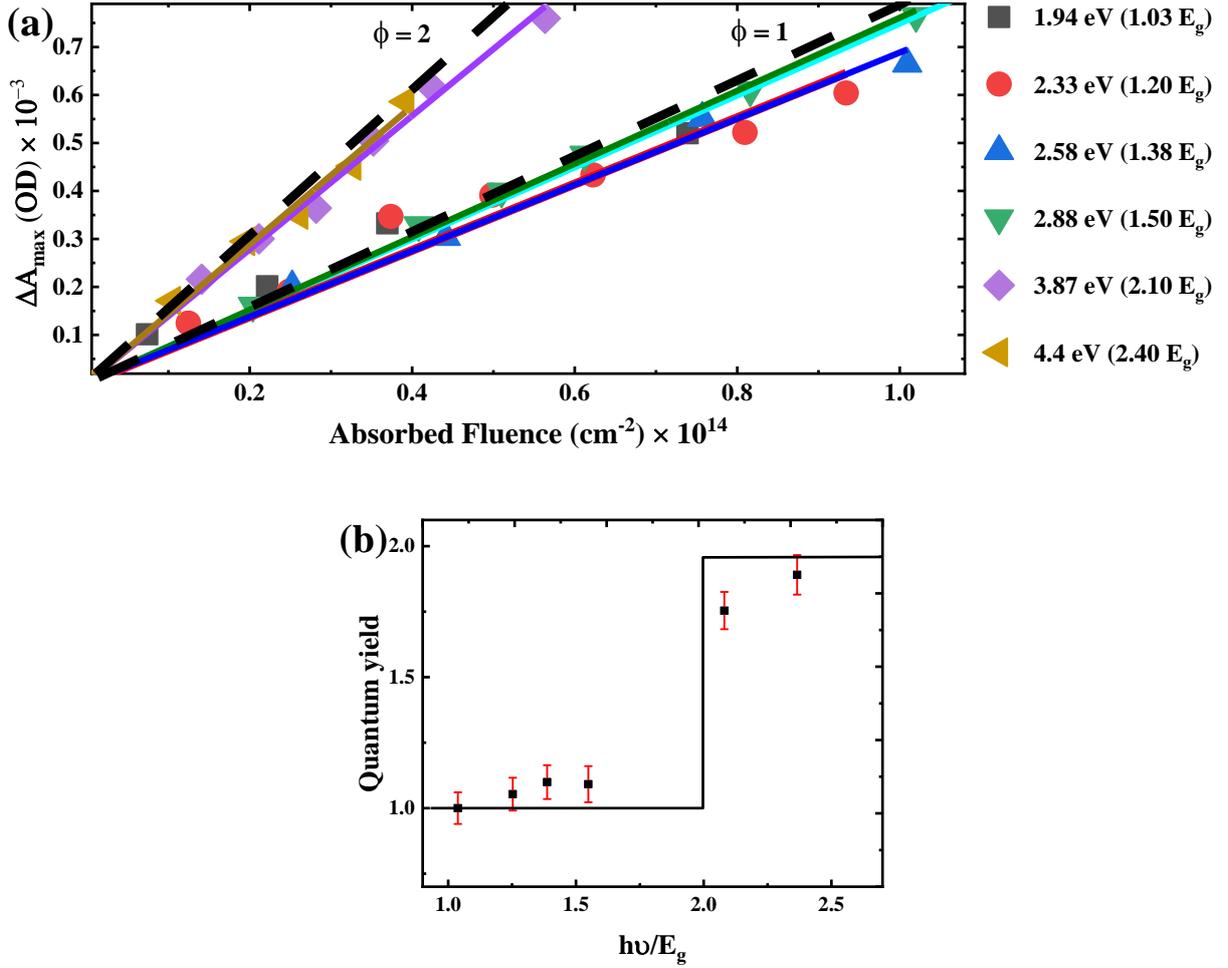

**Figure 5** (a) Variation of maximum PIB signal as a function of absorbed photon fluence at different pump energies. The solid lines represent the linear fit to experimental data points. The MEG quantum yields 1 and 2 are indicated by the dashed lines for reference. (b) Plot of MEG efficiency as a function of excitation energy normalized to the optical bandgap energy of the monolayer MoS₂. The data points correspond to the yield calculated from the A/B ratio (A and B are indicated in figure 3).

**Quantification of MEG quantum yield.** To determine the values of QY of MEG, we followed the procedure proposed by Weerd et al.[47] The maximum value of $\Delta A$ has been taken as a measure of the number of generated excitons (indicated by A in figure 3). As the excitons did not decay to zero within the measured time, we normalized the multiplicity of excitons to the B (PIB signal at the tail, figure 3) value. The experimental MEG QY at an energy greater than MEG threshold can be determined from the following relation

$$\emptyset(E_{at}) = \frac{A(E_{at})/B(E_{at})}{A(E_{bt})/B(E_{bt})} \qquad\qquad 4$$

where $E_{at}$ and $E_{bt}$ represent excitation energy above and below MEG threshold, respectively. The MEG QYs obtained from equation 4 are plotted as a function of pump photon energy normalized with the optical bandgap energy of $MoS_2$ in figure 5(b). A step-like increase in MEG QY after threshold energy (i.e., above $2E_g$) can be clearly seen. To estimate the MEG efficiency, we used the following model [11]

$$\emptyset = \eta_{cm} \left( \frac{h\nu}{E_g} - 1 \right) \qquad\qquad 5$$

where $h\nu$ is the excitation energy and $\eta_{cm}$ is the efficiency of MEG. We fitted the QYs obtained from equation 4 with the above-mentioned model (supplementary figure S5), which yields MEG efficiency of 86%. The observed large value of MEG efficiency infers that a major portion of excess photon energy is converted in e-h pair via MEG. Our results suggest that the multiple exciton generation in monolayer $MoS_2$ is sufficient to highlight its potential in solar conversion and harvesting applications.

**Conclusions**

By probing the carrier dynamics at various pump energies, we have demonstrated efficient MEG in monolayer $MoS_2$. An additional component in the decay dynamics of photogenerated carriers has been observed due to Auger recombination of multiple excitons which are formed at excitation energies higher than twice the optical bandgap energy. The longer build-up time of the PIB signal corresponding to above MEG-threshold excitation suggests that MEG process is not instantaneous. The generation of additional excitons are delayed because they are formed via relaxation of the photogenerated hot excitons. The threshold energy for the MEG is found to be as low as double the optical bandgap energy and MEG efficiency can goes up to ~86%. Our results provide global insight into the MEG process in monolayer $MoS_2$ and carrier multiplication in 2D $MoS_2$ makes it a promising material for highly efficient harvesting of light energy into electricity, enabling development of 2D TMD based high-efficiency optoelectronic devices.

**Experimental method**

**Sample preparation.** Monolayer $MoS_2$ flakes were synthesized on a sapphire substrate by chemical vapor deposition (CVD) technique in a one-inch horizontal hot tube furnace. Molybdenum trioxide powder ($MoO_3$) and sulphur powder were used as the growth reactant. The transition metal source and sapphire substrate were placed at the central high-temperature heating zone, while the sulphur powder was placed upstream and heated-up using a heating belt. The monolayer $MoS_2$ was grown at 850°C in the centre of the heating zone, and the sulphur temperature was kept at 135°C. A mixed gas of $H_2$/Ar with flow rate of 8/80 sccm was used as the carrier gas under a base pressure of 80 torrs during the growth process.

**Absorption spectroscopy.** A Shimadzu UV-2450 spectrometer (Agilent Technologies, USA) was used to measure the extinction spectrum within an excitation range of 1.7 eV to 4.96 eV. The measurement was performed on the sapphire substrate at room temperature.

**Raman and PL spectroscopy**. A, Horiba Jobin-Yvon LabRAM HR evolution Raman spectrometer was used for recording Raman and photoluminescence (PL) spectra in the back-scattering configuration under an optical microscope at room temperature. The sample was excited by a laser of wavelength 532 nm through a 100X objective lens. A monochromator analyzed the signals using a grating with 600 line/mm and 1800 line/mm for PL and Raman measurements, respectively. A Peltier cooled charge-coupled device detector in the back-scattering configuration was used to detect the signal.

**Femtosecond transient absorption (TA) spectroscopy.** A femtosecond Ti:Sapphire amplifier (pulse width < 35 fs and wavelength ~ 800 nm) was employed in the ultrafast Transient Absorption (TA) measuring system [58]. The pump and probe pulses were generated from the amplifier (Spitfire Ace, Spectra physics) by splitting the output into two components. The pump and probe pulses of different energies (used in single kinetics measurements) were obtained from the nonlinear optical parametric amplifier (TOPAS). In broadband TA measurement, the probe beam was a white-light continuum (WLC) generated by focusing a small fraction of 800 nm light (from Spitefire Ace) on a sapphire crystal. The 800 nm beam was adjusted using an iris and neutral density (ND) filter to obtain a stable and continuous white light probe. A delay stage in the probe path, run by a stepper motor, controls the time delay between probe and pump beams. The probe beam was split into the reference and the sample beams. These two beams were detected separately to eliminate the unsolicited noises. The probe beam was focused by a concave mirror with a hole at the center to allow the overlapping of the pump and probe beams at focus. The absorbance of the probe beam was detected under the conditions with and without pump, with the help of a mechanical chopper of frequency 500 Hz. A Berek compensator was used to control the polarization of the pump beam. The probe beam polarization was set at the 'magic angle' with the pump beam. TA spectra were recorded by dispersing the beam with a grating spectrograph (Acton Spectra Pro SP 2358) followed by a CCD array. Two photodiodes having variable gain were used to record TA kinetics. TA decay traces were fitted with multi-exponential functions by taking the instrument response function into account.

## ACKNOWLEDGMENTS

AS, DK, and SKP are thankful to the Advanced Materials Research Centre (AMRC) of IIT Mandi for providing experimental facilities. SKP acknowledges the financial support from Science and Engineering Research Board (SERB), Government of India (Grant No. CRG/2018/003045).

# Efficient Multiple Exciton Generation in Monolayer MoS$_2$

*Ashish Soni [1,2], Dushyant Kushavah [1,2], Li-Syuan Lu [3], Wen-Hao Chang[3,4] and Suman Kalyan Pal [1,2]\**

[1] School of Physical Sciences, Indian Institute of Technology Mandi, Kamand, Mandi 175005 Himachal Pradesh, India

[2] Advanced Materials Research Centre, Indian Institute of Technology Mandi, Kamand, Mandi 175005 Himachal Pradesh, India

[3] Department of Electrophysics, National Yang Ming Chiao Tung University, Hsinchu 30010, Taiwan

[4] Research Center for Applied Sciences, Academia Sinica, Nankang, Taipei 11529, Taiwan

[*] Email: suman@iitmandi.ac.in

**Correction of absorption spectrum.** Thin film samples scatter the incoming light elastically, known as Rayleigh scattering. This affects the absorption spectrum since the scattered light does not reach the detector of the absorption spectrometer and, thus, counted as absorbed light by the instrument. The measured absorption spectrum of our sample (Fig. S1) exhibits very high baseline absorption which could be attributed to Rayleigh scattering by $MoS_2$ sample.

The Raleigh scattering is proportional to $\lambda^{-4}$ i.e., scattering is more intense towards the blue side of the spectrum.[1, 2] Absorption due to pure Rayleigh scattering can be expressed as

$$A = log(I_0/I) + A_0$$

$$= \log \frac{I_0}{I_0 - I_{scatter}} + A_0$$

$$= \log \frac{1}{1 - c \times \lambda^{-4}} + A_0 \qquad\qquad 1$$

where $I_0$ and $I$ are intensity of incident and transmitted lights, respectively, $A_0$ and $c$ are constants.

To remove the effect of scattering from the absorption spectrum of our sample, we followed the below-mentioned steps.

1. First, we chose a data range in the high wavelength side of the spectrum as marked by a dotted circle in figure S1. In this region, the absorption due to the sample is expected to be 0.

2. Next, we fitted the selected interval with the function given in equation 1 and find the values of $A_0$ and $c$. By using the values of these constant parameters, we determined baseline (blue line in Fig. S1) for the complete wavelength range (300-800 nm).

3.  The calculated baseline was subtracted from the measured absorption spectrum to get corrected spectrum (black curve, Fig. S1).

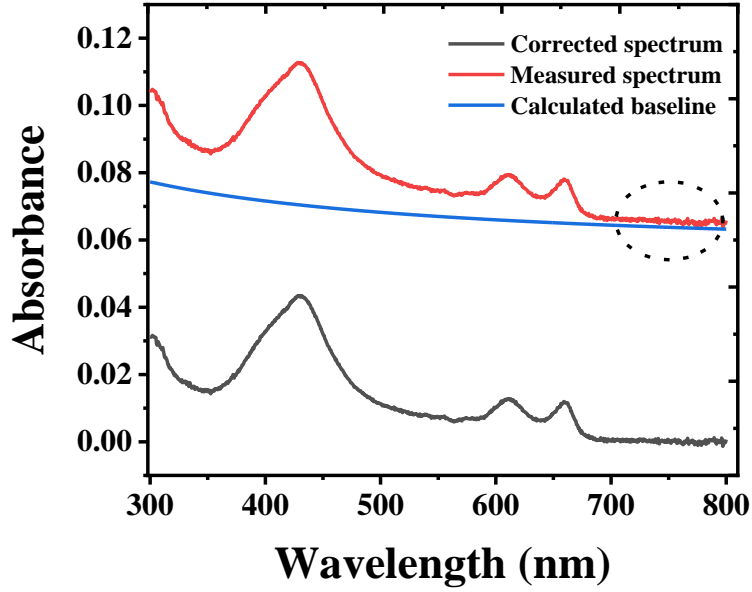

**Figure S1** Absorption spectrum of monolayer $MoS_2$ sample without (red) and with (black) baseline correction considering Rayleigh scattering.

**Determination of absorbed photon density.** We determined the absorbed photon fluence using the value of pump fluence and the fraction of absorbed photons,

$$Absorbed\ photon\ fluence = Fraction\ of\ photon\ absorbed\ \times\ Pump\ fluence \qquad 2$$

We calculated the pump fluence (pump photon density) of the laser beam falling on the sample using the relation,

$$Pump\ fluence\ (cm^{-2}) = \frac{X \times 10^{-6}\ joule/sec}{500\ Hz \times E_p\ (joule) \times A(cm^2)} \qquad 3$$

where $X$ is the power of laser beam at the sample in microwatt, $E_p \left( = \frac{hc}{\lambda} \right)$ is the energy of a single photon, and $A$ is the cross-section area of the pump beam.

To obtain the fraction of absorbed fluence, we utilized the Beer-Lambert law

$$A = \log_{10}\left(\frac{I_0}{I}\right) \qquad 4$$

The fraction of absorbed photons

$$\frac{(I_0 - I)}{I_0} = 1 - 10^{-A} \qquad 5$$

where $A$ is the steady state absorbance.

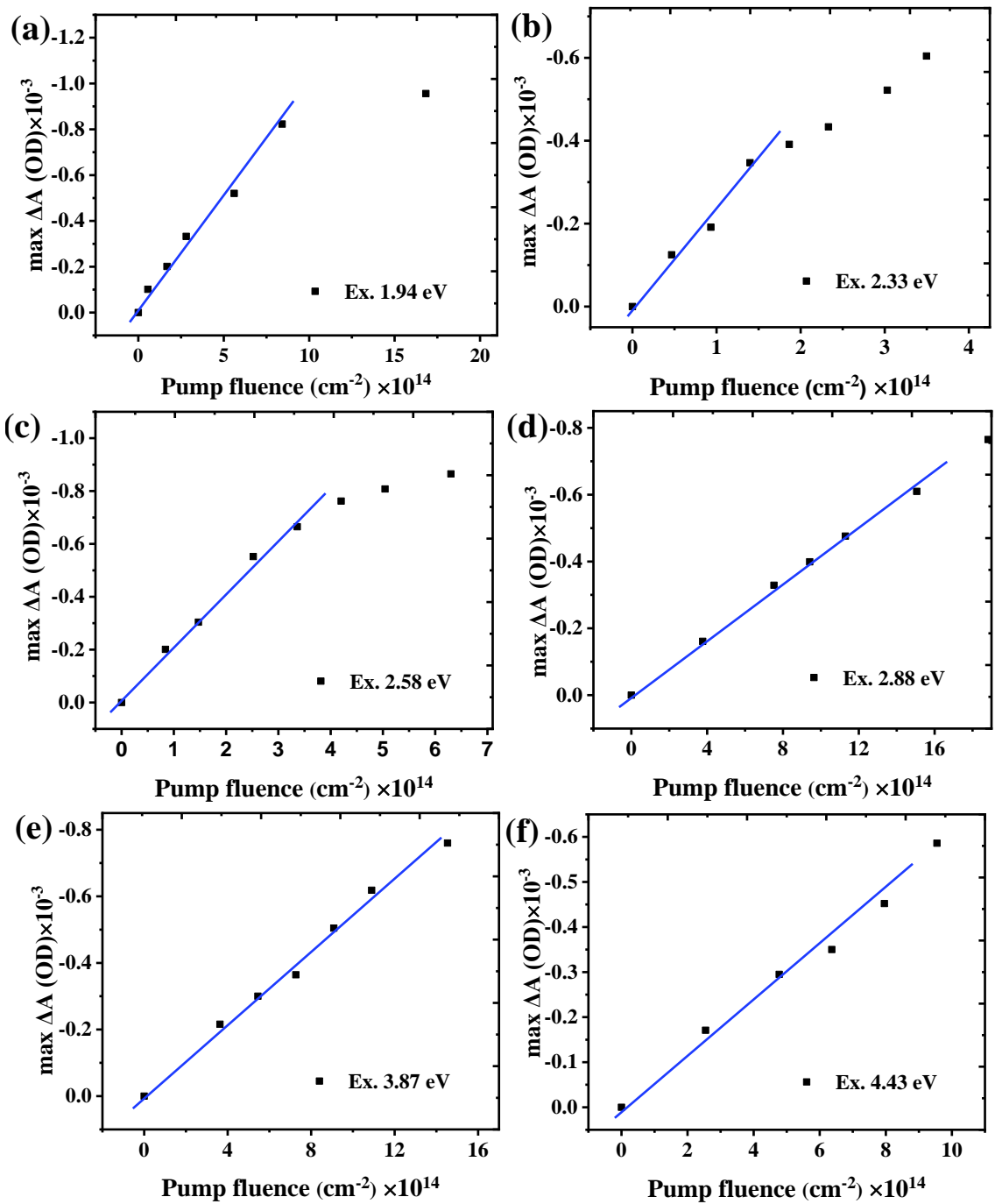

**Figure S2** Linear relation between pump fluence and maximum PIB signal at pump photon energies 1.94, 2.33, 2.58, 2.88, 3.87 and 4.4 eV.

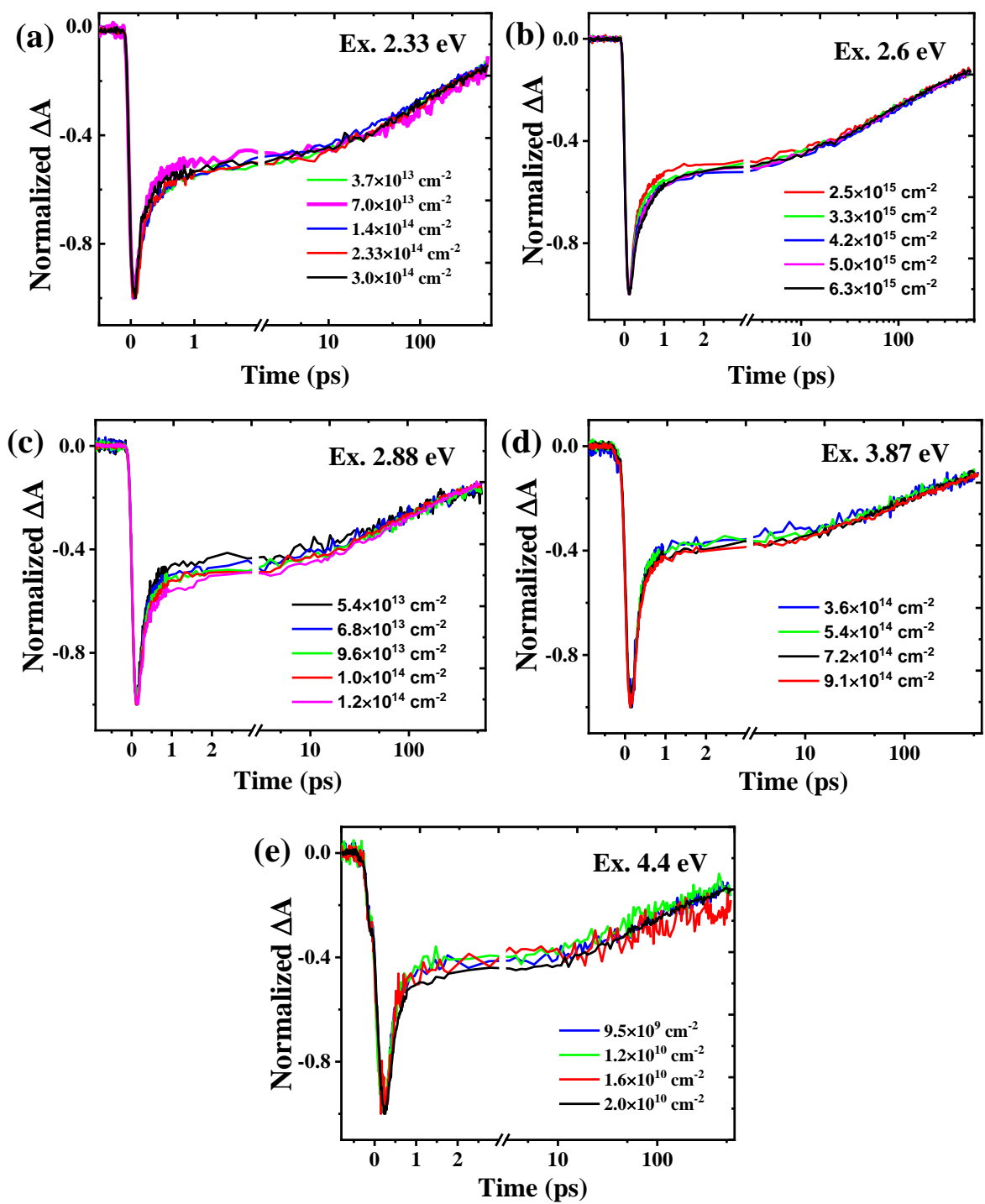

**Figure S3** Normalized photoinduced bleach kinetics measured at different excitation fluences and excitation energies. The decays at different pump fluences follow almost the same kinetics which confirms that our measurements were done in the linear regime.

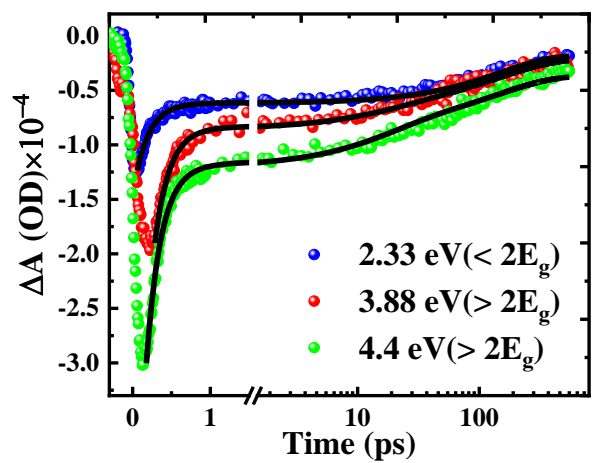

**Figure S4** Multiexponential global fitting of PIB kinetics measured at excitation photon energy 2.33 eV (<2E$_g$), 3.87 eV (>2E$_g$), and 4.4 eV (>2E$_g$) with a fixed photon fluence.

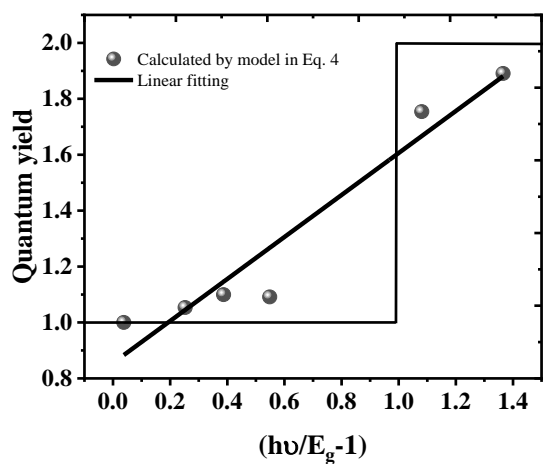

**Figure S5** Quantum yields of MEG at different pump photon energies calculated by the model described by equation 4 in the main text. The solid line is the result of the fitting of experimental data with equation 5 in the main text. The slope of the straight line yields the efficiency of MEG ~86%.

**Table S1** Parameters obtained from multiexponential global fitting of PIB kinetics measured at excitation photon energy 2.33 eV with increasing pump fluence

| Pump Fluence ($cm^{-2}$) | $T_1$ (ps) | $A_1$ (%) | $T_2$ (ps) | $A_2$ (%) | $T_3$ (ps) | $A_3$ (%) |
|---|---|---|---|---|---|---|
| $3.7 \times 10^{13}$ | 0.20 | 68 | - | - | 150 | 32 |
| $7.0 \times 10^{13}$ | | 67 | | | | 33 |
| $1.4 \times 10^{14}$ | | 66 | 13 | 5 | | 28 |
| $1.8 \times 10^{14}$ | | 62 | | 6 | | 33 |
| $2.3 \times 10^{14}$ | | 60 | | 6 | | 34 |

**Table S2** Parameters obtained from multiexponential global fitting of PIB kinetics measured at excitation photon energy 3.87 eV with increasing pump fluence

| Pump Fluence ($cm^{-2}$) | $T_1$ (ps) | $A_1$ (%) | $T_2$ (ps) | $A_2$ (%) | $T_3$ (ps) | $A_3$ (%) |
|---|---|---|---|---|---|---|
| $3.6 \times 10^{14}$ | 0.23 | 82 | 14 | 5 | 150 | 13 |
| $5.4 \times 10^{14}$ | | 81 | | 5 | | 14 |
| $7.2 \times 10^{14}$ | | 80 | | 6 | | 14 |
| $9.0 \times 10^{14}$ | | 79 | | 7 | | 14 |
| $1.0 \times 10^{15}$ | | 78 | | 8 | | 14 |

**Table S3** Parameters obtained from multiexponential global fitting of PIB kinetics measured at excitation photon energies 2.33, 3.87 and 4.4 eV

| Pump energy (eV) | $T_1$ (ps) | $A_1$ (%) | $T_2$ (ps) | $A_2$ (%) | $T_3$ (ps) | $A_3$ (%) |
|---|---|---|---|---|---|---|
| 2.33 | 0.19 | 66 | 17 | - | 150 | 34 |
| 3.87 | | 84 | | 7 | | 9 |
| 4.40 | | 86 | | 6 | | 8 |